# Effect of trip attributes on ridehailing driver trip request acceptance[1]

Yuanjie Tu[a*], Moein Khaloei[a], Nazmul Arefin Khan[b], Don MacKenzie[a]

[a] *Department of Civil and Environmental Engineering, University of Washington, Seattle, US*

[b] *Argonne National Laboratory, Lemont, US*

*Corresponding author Email: yuanjt2@uw.edu

---

[1] Paper in print at *Journal of Sustainable Transportation*.

# Effect of trip attributes on ridehailing driver trip request acceptance

A generalized additive mixed model was estimated to investigate the factors that impact ridehailing drivers' trip request acceptance choices, relying on 200 responses from a stated preference survey in Seattle, US. The main findings are: (1) ridehailing drivers are less likely to accept trips with longer pick-up times; the relationship is linear for short trips when the trip is below 45 minutes, while nonlinear otherwise; (2) trips over 45 minutes, surge price, and higher passenger ratings increase the probability of trip request acceptance; (3) ridehailing drivers are less likely to accept shared trips than solo ones; (4) the effects of ridehailing drivers' socio-demographics and employment status are statistically non-significant. Several policy recommendations were proposed to promote trip request acceptance based on ridehailing drivers' willingness to accept compensation for undesired trip features. The findings could be useful for transportation agencies to improve ridehailing service efficiency, better fulfill urban mobility needs, and reduce environmental burden.

**Introduction**

Trip request acceptance by ridehailing drivers is critical to policymakers, planners, engineers, and the public: higher trip request acceptance rates contribute to both social and environmental sustainability. When trip requests are sent first to the nearest driver, higher trip request acceptance rates not only shortens passenger waiting times, but might also reduce the disparities of passenger waiting times across different races (Ge, Knittel, MacKenzie, & Zoepf, 2020) and neighbourhoods. Higher trip request acceptance rates also reduce deadheading, therefore save total vehicle miles traveled (VMT), which helps reduce greenhouse gas emissions. However, since drivers are independent contractors, transportation network companies (TNCs) that match drivers with passengers are limited in their ability to dictate drivers' actions towards trip requests (Wentrup, Nakamura, & Ström, 2018). Therefore, it is necessary to understand disaggregate driver trip request acceptance behaviors and preferences if policymakers are to develop interventions that effectively nudge drivers' choices in desired directions.

This study modeled the relationship between trip features and drivers' trip request acceptance using data from a stated choice experiment. Based on our findings, we have further provided policy recommendations for promoting trip request acceptance rates by calculating drivers' willingness to accept compensation for undesired trip features.

**Literature review**

*The importance of driver behavior*

Transportation scholars have long recognized the importance of travelers' behaviors and the need to understand their preferences and motives if we are to forecast travel demand or develop effective demand management policies. Accordingly, a large body of TNC-related research has focused on estimating traveler behavior and choices towards ridehailing services (Brown, 2020; Payyanadan & Lee, 2018; Wu & MacKenzie, 2021). Another considerable body of research has examined system-level impacts of TNCs such as congestion and public transit ridership (Diao, Kong, & Zhao, 2021; Erhardt et al., 2021; Graehler, Mucci, & Erhardt, 2019), discrimination effects (Bokányi & Hannák, 2020; Ge et al., 2020) and environmental impacts (Barnes, Guo, & Borgo, 2020; Ward et al., 2019). Others have examined the reverse: the effects of system conditions on TNC trip fare, duration, and pick-up waiting time (Shokoohyar, Sobhani, & Sobhani, 2020). A number of studies have investigated the distribution of access and waiting times across neighborhoods (Hassanpour, Bigazzi, & MacKenzie, 2021; Shokoohyar, Sobhani, & Ramezanpour, 2020; Wang & Mu, 2018; Hughes & MacKenzie, 2016). Although such outcomes are generally affected by driver behaviors, prior work has mainly examined indirect, system-level indicators related to ridehailing services.

*Prior studies of driver behavior*

Understanding ridehailing driver behaviors and their effects on platform efficiency, equity, and the environment is of great importance (Shokoohyar, 2018). A handful of studies have qualitatively analyzed and quantitatively modeled the impacts of trip features on driver response to a request (Ashkrof et al., 2020, 2021; Morris et al., 2020; Xu, Sun, Liu, & Wang, 2018).

Xu et al. (2018) investigated factors that impact ridehailing driver trip request acceptance using data from a ridehailing service provider in Beijing, China. This study discovered that ridehailing drivers are more likely to accept long trips and trips with surging prices. They also found that the number of trip requests received, and the number of trips accepted per driver have little impact on trip request acceptance.

Ashkrof et al. (2020) studied driver behavior and preferences on trip request acceptance, working shift, and relocation choices. They collected data in a series of focus groups with 16 Uber drivers in the Netherlands. They found that pick-up locations, distance and time to the pick-up point, passenger ratings, surge pricing, long distance rides, destination prediction, driver's experience, and cancellation criteria may be related to a driver's decision to accept or reject a trip request. Drivers stated that they would be more likely to reject trips in risky locations, trips that require long pickup time and distance, and trips with low passenger ratings. Experienced drivers were more selective about accepting rides.

In a grey literature study, Ashkrof et al. (2021) examined trip request acceptance behavior by collecting a stated preference survey of ridehailing drivers in the United States (US) (752 responses) and in the Netherlands (68 responses). By using a discrete choice model, they found that employment status of the driver, amount of time driving, working shift, travel time to the pick-up point, and surge price were important predictors of trip request acceptance.

A related study on driver attitudes towards shared rides surveyed 309 ridehailing drivers across the US (Morris et al., 2020). Results suggested that in general, drivers dislike shared rides compared with solo rides. They discovered that ridehailing drivers are generally dissatisfied with shared rides because they are less cost-effective. That study does not link drivers' dislike towards shared rides to trip request acceptance behaviors, but it does help identify a potential factor that may affect trip request acceptance.

Additional evidence in online forums and blogs shows some drivers expressing concerns about passengers with low ratings, and indicating they are less likely to accept a trip request with a low passenger rating (Bowman, 2019; Helling, 2021; RideGuru, 2018).

*Contribution of This Study*

There are two limitations in current literature. First, no studies have comprehensively considered the effects of trip features, socio-demographics, and employment status on driver behavior. Second, no studies have explored the non-linear relationships between trip request acceptance and trip attributes.

This study advances the current research area by quantifying the linear and nonlinear impacts of trip features on drivers' trip request acceptance. To the authors' knowledge, this is the first study applying a generalized additive mixed model (GAMM) to examine these dependencies alongside the effects of drivers' socio-demographic and employment characteristics. The findings and proposed policy implications can also be used to improve ridehailing service efficiency, better fulfil urban mobility needs, narrow the mobility disparities across different races and neighborhoods, and reduce negative environmental impacts.

**Data collection**

Our primary data source is a stated preference survey completed by 200 ridehailing drivers in the Seattle metropolitan area. Using a stated preference survey enables us to investigate the causal relationship between trip features and trip request acceptance rate. The inclusion of an experimental design in the survey allows us to control endogeneity, a common problem that can lead to omitted variable bias when using non-experimental data (Abdallah, Goergen, & O'Sullivan, 2015).

The study was identified as human subjects research that qualifies for exempt status from Institutional Review Board (IRB) approval by Human Subjects Division, University of Washington (IRB ID: STUDY00013095). The survey was conducted from August 11 to September 9, 2021 in Seattle, USA. We trialed online and in-person data collection approaches. An in-person approach was finally adopted because (1) the online response rate was low and (2) many online respondents appeared not to be real ridehailing drivers, based on screening and quality check questions in the survey. More details on performance of different data collection approaches can be found in (Tu, Jabbari, Khaloei, & MacKenzie, 2021).

Participants were recruited in the SeaTac airport ridehailing driver waiting area and assisted in completing the survey on an iPad. Respondents were compensated with a $15 Amazon gift card.

Anecdotally, ridehailing drivers do not seem to trust surveys and interviews. The most frequently asked questions when administering the survey are "what is the purpose of the study?" and "do you work for Lyft or Uber?" Their major concern was that the responses would be used against them by ridehailing companies. This echoes existing evidence that drivers and companies have a tense relationship (Ashkrof et al., 2020). A total of 200 responses were collected from an estimated 250 drivers who were invited to participate.

*Survey design*

Our team conducted a series of pilot interviews with ridehailing drivers before designing the questionnaire, to select the most important variables and identify terminology familiar to the drivers. For example, the questionnaire used the term "ride-sharing" rather than "ridehailing" because "ride-sharing" is widely used among the drivers. The questionnaire contained three sections: basic driving information, choice experiments, and background information.

*Basic driving information*

The respondents were asked to provide information on which ridehailing companies they are driving for, how long they have been a ridehailing driver, whether they have another job besides ridehailing driver, the time they generally start and stop driving each day, target working hours and earnings per day, and the numbers of trip requests they (1) received and (2) rejected in the past week.

*Choice experiments*

To investigate the causal relationship between the trip features and trip request acceptance, a blocked factorial design was adopted. The experimental design was conducted as follows. First, 5 variables were chosen as experimental variables, including ridehailing demand, passenger pick-up time, passenger ratings, solo/shared trip, and whether the trip is over 45 minutes. These variables included the information commonly provided to drivers when they receive a trip request. Variable descriptions are shown in Table 1. Second, levels were generated for each experimental variable (Table 2). Third, the full matrix of combinations of the variable levels was generated. Next, we created blocks of 6 combinations (choice scenarios) for each respondent. Each respondent was asked to make choices for 6 choice scenarios. In each scenario, she/he was asked to choose to "accept" or "reject" a trip request based on different combinations of experimental variable levels. Figure 1 shows an example of driver trip request acceptance experiment.

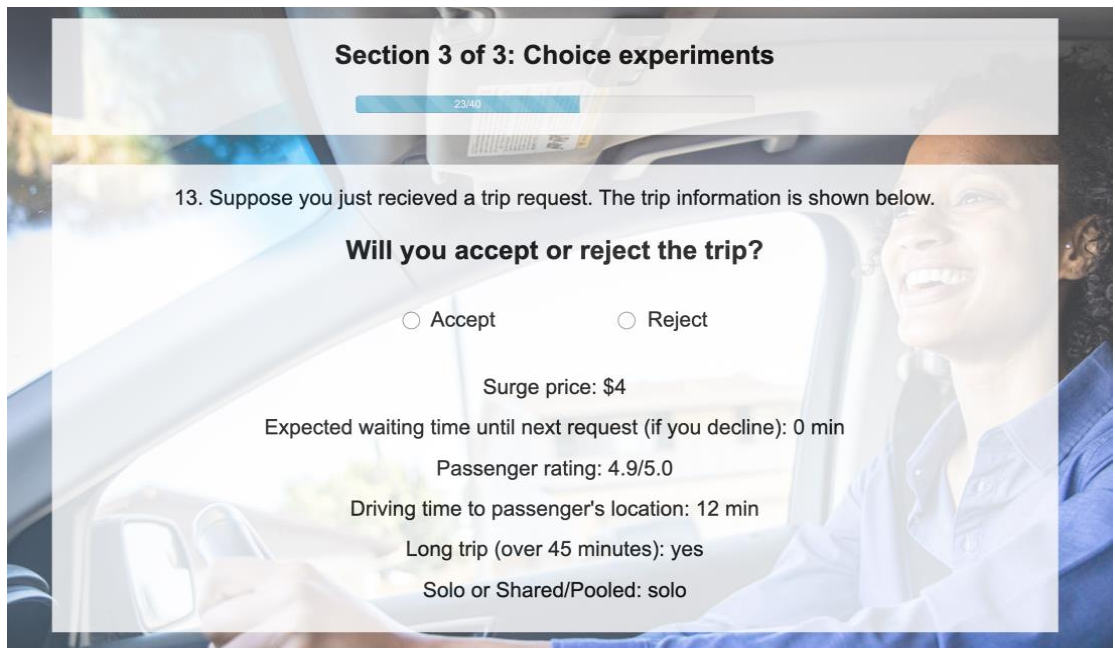

Figure 1. An example of trip request acceptance

*Background information*

In this section, respondents were asked questions about their socio-demographics including age, gender, race, whether born in the US, whether a student, household size, individual and household income, education level, job status (whether part time, whether they have another job), and subjective health status.

*Data quality*

To improve data quality, multiple screening and quality check questions were adopted as follows. First, before entering the choice scenario questions, survey respondents were asked two comprehension check questions. For each comprehension check, the respondents were asked to choose the correct meaning of a highlighted number in a choice scenario. Every respondent had the chance to answer each of the comprehension questions twice. They were only allowed to proceed if the answers were correct, otherwise the survey was terminated. Second, one question was included to check respondents' attention. The respondents were told "this is an attention check" and were asked to choose the neighborhood with given surge price, waiting time, and travel time in the middle of choice scenario questions. Finally, a logic check was applied in the survey. In the background information section, respondents were asked to provide their annual income (both household and individual). Each respondent's annual household income should be equal or larger than their individual income. Responses failing either the attention or logic check, were removed.

For modeling purposes, we also removed respondents who have null inputs or choosing "prefer not to answer" in independent variables mentioned in Section 3. Respondents who completed 0 trips in the prior week were also removed.

After data quality check and data cleaning, a total of 181 respondents and 1,085 choice responses were retained for analysis.

**Table 1.** Experimental variables

| Attributes | Definition |
|---|---|
| Ridehailing demand | Represents the demand for ridehailing services. Measured jointly by *surge price ($)* and *average waiting time for a trip request if they decline the current one (min)*. Higher demand areas have lower waiting times and higher surge price and vice versa. |
| Passenger pick-up time | The time to drive from the current location to the pickup location (minutes). |
| Passenger ratings | Average rating the passenger received from drivers, ranging from 4.1 to 4.9. We chose 4.1 as the lower bound because we learned from driver interviews that very few passengers have ratings below 4 stars. |
| Solo/shared trip | Whether the trip is a solo or shared trip. A shared trip means picking up unrelated passengers or groups of passengers with different origins and destinations. A solo trip means an individual or group traveling together. |
| Long trip | If the trip is over 45 minutes, 1; otherwise, 0. We choose the 45-minute threshold because the pilot driver interviewees mentioned ridehailing companies provide 45 minutes as a threshold for "long trip". |

**Table 2.** Experimental attribute levels

| Attributes | | | | | | | | | | |
|---|---|---|---|---|---|---|---|---|---|---|
| Ridehailing demand | Surge price ($) | 16 | 12 | 8 | 4 | 0 | 0 | 0 | 0 | 0 |
| | Average waiting time (min) | 0 | 0 | 0 | 0 | 0 | 8 | 16 | 24 | 30 |
| Passenger pick-up time (min) | | 4 | | 8 | | 12 | | 16 | | 20 |
| Passenger ratings | | 4.1 | | 4.3 | | 4.5 | | 4.7 | | 4.9 |
| Solo/shared trip | | Solo | | | | | Shared | | | |
| Long trip | | Yes, long trip | | | | | No, not a long trip | | | |

**Methods**

To understand the relationship between trip features and ridehailing drivers' behavior towards trip request, a generalized additive mixed model (GAMM) was adopted.

The dependent variable is a binary variable, meaning that a respondent can "accept" or "reject" the trip request based on the given choice scenarios. Independent variables include driving habits (trip rejection rate), trip features (e.g., pick-up time, long trip, surge price), socio-demographics (e.g., age, gender) and employment characteristics (e.g., whether they have another job besides ridehailing driver). To account for multicollinearity, variables with variance inflation factors (VIF) larger than 5 were excluded. The variable descriptions are shown in Table 3.

A GAMM is a semi-parametric statistical modeling approach widely used in detecting nonlinear relationships (Wood, 2017). The nonlinear effects represented by nonparametric functions shown in Equation (1) are represented by smooth splines and estimated by restricted maximum likelihood. In addition to accommodating nonlinear effects, a GAMM includes a random effect for each individual driver, which corrects for the non-independence of repeated observations of the same respondent. The R package "mgcv" was used to estimate the model (Wood, 2021). Definitions of the variables are shown in Table 4.

$$g(\mu_{ik}) = \alpha + \beta X_{ik} + \gamma Z_i + f(r_{ik}, l_{ik}) + f(wr_i) + b_i \qquad (1)$$

**Table 3.** Variable descriptions

| Variable | Definition |
|---|---|
| *Dependent variable* | |
| Response to a request | If the respondent accepts the trip request, 1, else, 0 |
| *Driving habits* | |
| Trip rejection rate | Number of trip requests rejected divided by number of total trip requests received last week, in % |
| *Trip features* | |
| Pick-up time | Driving time to pick up a passenger, in minutes |
| Long trip | If the trip is longer than 45 minutes (does not include pick-up time), 1; else, 0 |
| Average trip waiting time | Expected waiting time until next request if the respondent rejects the current one, in minutes |
| Surge price | An additional surge amount to the trip fare, in dollars |
| Passenger ratings | The average of the ratings that a passenger has received from drivers |
| Shared/pooled ride | The trip is shared with other passengers who are heading in the same direction. If it's a shared trip, 1; else, 0 |
| *Socio-demographics* | |
| Age | The age of the respondent.<br>18 – 39: 0<br>40 – 64: 1<br>>=65: 2 |
| Gender | If 1, female; 0, male |

| Household size | The number of people (including the respondent) in the household. |
|---|---|
| City median household income | If household income equal or higher than Seattle median household income in 2019 ($102,500) (Balk, 2020), 1; else, 0. |
| Education level | The education level of a respondent<br>0: less than high school<br>1: high school or GED<br>2: college, including some college, professional degree, associate degree and bachelor's degree<br>3: graduate school or above, including master's degree and doctoral degree |
| *Employment characteristics* | |
| Uber | If Uber driver, 1; else, 0 |
| Lyft | If Lyft driver, 1; else, 0 |
| Another job | Whether the respondent has another job other than the ridehailing driver<br>0: has another job, full-time (35+ hours/week)<br>1: has another job, part-time (fewer than 35 hours/week)<br>2: doesn't have another job |

**Table 4.** Variable definitions.

| Notation | Descriptions |
|---|---|
| $g(\cdot)$ | a logit function |
| $\mu_{ik}$ | $E(y_{ik})$, where $y_{ik}$ response to a request of driver $i$ and choice scenario $k$ |
| $\alpha$ | the constant |
| $X_{ik}$ | a vector of attributes of choice situation $k$ faced by driver $i$ |
| $Z_i$ | a vector of observable attributes of driver $i$ |
| $f(\cdot)$ | a smooth function (we use low-rank thin plate splines in this study). |
| $r_{ik}$ | pick-up time |
| $l_{ik}$ | a dummy variable representing whether the trip presented to driver $i$ in scenario $k$ is a long trip |
| $f(r_{ik}, l_{ik})$ | nonlinear interactions between pick-up time and long trip. |
| $wr_i$ | the weekly trip rejection rate of driver $i$ |
| $f(wr_i)$ | a smooth function between trip rejection rate and trip request acceptance. In this study, we only included two nonlinear terms because other variables were found to have a linear relationship with $g(\mu_{ik})$. |
| $b_i$ | an individual-specific random component assumed to be distributed as $N\{0, \sigma^2\}$ where $\sigma$ is the variance component |

**Results**

*Descriptive analysis*

*Socio-demographics and employment characteristics*

The data characteristics are shown in Table 5. Compared to the general population, our sample of ridehailing drivers has some unique characteristics. They are overwhelmingly male (95%), African American (75%), and born outside the US (94%). Most of the respondents are working age, with only 4% being 65 years or older. In terms of education background, a vast majority of respondents – 46% and 44% graduated from high school and college, respectively. Only a small proportion of respondents have an education background less than high school or higher than college (graduate school).

Most respondents (75%) have a small household size (<= 4 members). 30% of respondents live by themselves. Most respondents (58%) do not have children, followed by 1~4 children. Over half of respondents (52%) have a household income between $35,000 and $74,999. 28% have a household income less than $35,000. Only 17% have a household income higher than $75,000.

*Response to a request*

Overall, respondents reported accepting most of the trip requests they receive. 32% respondents said they had accepted all trips in the past week, and the average rejection rate in the past week was 5.9% (Figure 2A). Only 10% respondents said they had rejected more than 15 trips in the prior week. In the stated choice scenarios, 24% of the hypothetical trip requests were rejected (Figure 2B).

Driver respondents mentioned four reasons for accepting trips in the pilot interviews and the survey comment area: (1) rejecting trips brings punishment –driver ratings could decrease, there could be up to a 15-minute timeout with no trips, and there is a risk of account deactivation if they reject too many trips; (2) their income mainly comes from driving for TNCs, and they would be unable to make profit if they reject a lot of trips; (3) they are given approximately only 5 seconds to make the decision, hence it is difficult to be thoughtful and selective; and (4) the information provided to drivers is limited. Important information such as trip destination, estimated distance and time, and trip fare are not shown to the drivers with the trip request, which makes it difficult to turn down a request.

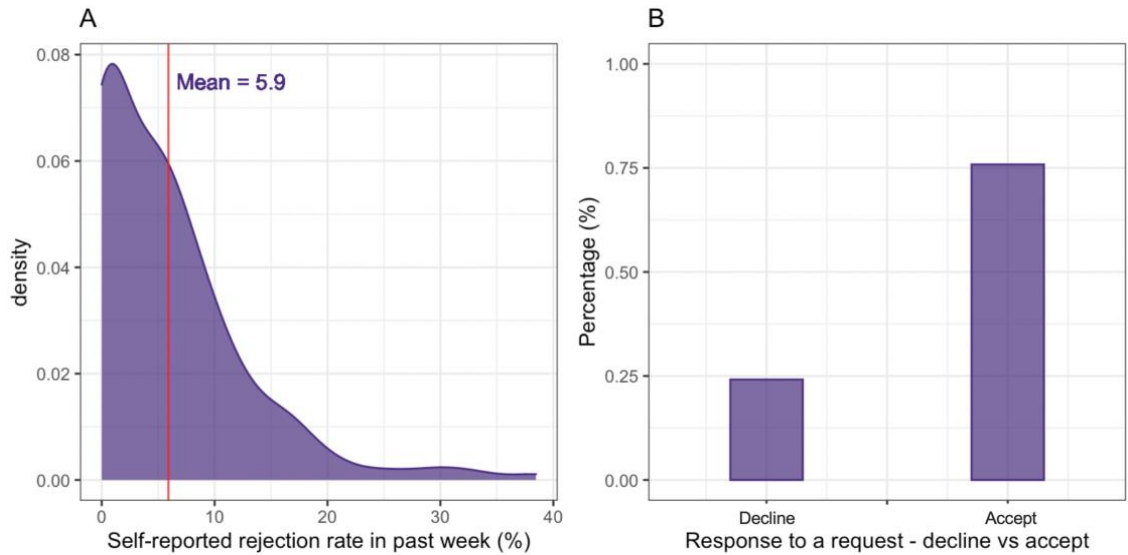

**Figure 2.** Distribution of response to a request

**Table 5.** Data characteristics (N = 181)

| Characteristic | This sample | Characteristics | This sample |
|---|---|---|---|
| *What is your gender?* | | *Are you currently a student?* | |
| Male | 96% | Yes, full time (35+ hours/week) | 3% |
| Female | 4% | Yes, part time (<35 hours/week) | 7% |
| *What is your race?* | | No, not a student | 90% |
| White | 7% | *Are you of Hispanic, Latino, or Spanish origin?* | |
| Asian | 11% | Yes, Hispanic origin | 2% |
| African American | 75% | No, non-Hispanic origin | 95% |
| Another | 5% | Prefer not to answer | 3% |
| Prefer not to answer | 2% | *Were you born in the United States?* | |
| *What is your age?* | | Yes, born in the US | 4% |
| 18-39 | 51% | No, born outside the US | 95% |
| 40-64 | 45% | Prefer not to answer | 1% |
| 65 and above | 4% | *Do you have another job besides ride-sharing driver?* | |
| *What is the highest degree or level of school you have completed?* | | Yes, full time (35+ hours/week) | 9% |
| Less than high school | 5% | Yes, part time (<35 hours/week) | 12% |
| High school | 46% | No, no another job | 79% |
| College | 44% | *Which category best describes your household income before taxes from the last calendar year?* | |
| Graduate school or higher | 5% | Less than $10,000 | 4% |
| *How long have you been an active ride-sharing driver? "Active" means the time between two rides should be no longer than a month.* | | $10,000 to $14,999 | 3% |
| Less than 6 months | 6% | $15,000 to $19,999 | 4% |
| 6 months - less than 1 year | 5% | $20,000 to $24,999 | 9% |
| 1 year - less than 1.5 years | 2% | $25,000 to $34,999 | 9% |
| 1.5 years - less than 2 years | 6% | $35,000 to $49,999 | 20% |
| 2 years or more | 80% | $50,000 to $74,999 | 33% |
| Prefer not to answer | 1% | $75,000 to $99,999 | 10% |
| *How many people live in your household including yourself?* | | $100,000 to $199,999 | 7% |
| 1 | 30% | $200,000 to $249,999 | 1% |
| 2 | 17% | *Which ride-sharing companies are you driving for?* | |
| 3 | 13% | Uber only | 14% |
| 4 | 15% | Lyft only | 35% |
| 5 | 11% | Uber and Lyft | 51% |
| 6 | 6% | *Within your household, how many are children under the age of 18?* | |
| >=7 | 8% | 0 | 58% |
| *How many trip requests did you complete in the past week?* | | 1 | 15% |
| 0~49 | 42% | 2 | 10% |
| 50~99 | 47% | 3 | 8% |
| 100~149 | 10% | 4 | 4% |
| >=150 | 1% | >=5 | 6% |

*Modeling results*

Table 6 reports the results of two models: a baseline model only linear main effects, and a full model that includes smoothed terms and interactions. The modeling results are shown in Table 6. The model results include two sections: model performance and coefficients. In terms of the model performance, the final log likelihood of the baseline model and the full model are -370 and -278, respectively. 38.3% and 53.5% deviance was explained by the baseline model and full model, respectively. Since the model results show that the full model has a better performance, the following discussion focuses on the results of the full model.

The coefficient section is composed of two parts: parametric coefficients and smooth terms. The parametric coefficient section shows the linear terms and their significance. The smooth term section exhibits the nonlinear terms and their estimated degrees of freedom. A variable with an estimated degree of freedom of 1 indicates a linear relationship, while estimated degrees of freedom significantly higher than 1 indicates nonlinearity.

*Trip features*

(1) Nonlinear terms: pick-up time

Figure 3 shows the nonlinear effects of pick-up time and its interactions with long trip from the GAMM. The horizontal axis represents pick-up time, while the vertical axis represents the effects on the log odds of accepting a trip versus rejecting a trip, everything else being equal. In both conditions, a driver is less likely to accept a trip request if s/he needs to drive a long time to pick up the passenger. A possible explanation is that trips with long pickups are commonly not cost-effective since drivers must cover the cost (e.g., the opportunity cost of getting another ride, fuel, mileage) by him/herself. This echoes the finding of another study, which found that a few drivers in a focus group said they tend to not take trips with long pickups (Ashkrof et al., 2020).

However, the interaction between pick-up time and long trip requests (over 45 minutes) reveals richer information. The relationship between pick-up time and trip request acceptance is linear when the trip is short. For short trips, the log odds of trip request acceptance decline linearly with pick-up time, even for short values of pick-up time. For long trips, however, the story is quite different: drivers appear to be insensitive to changes in pick-up time up to about 12 minutes. Once pick-up time exceeds 12 minutes, a driver's odds of accepting a long trip begin to decline.

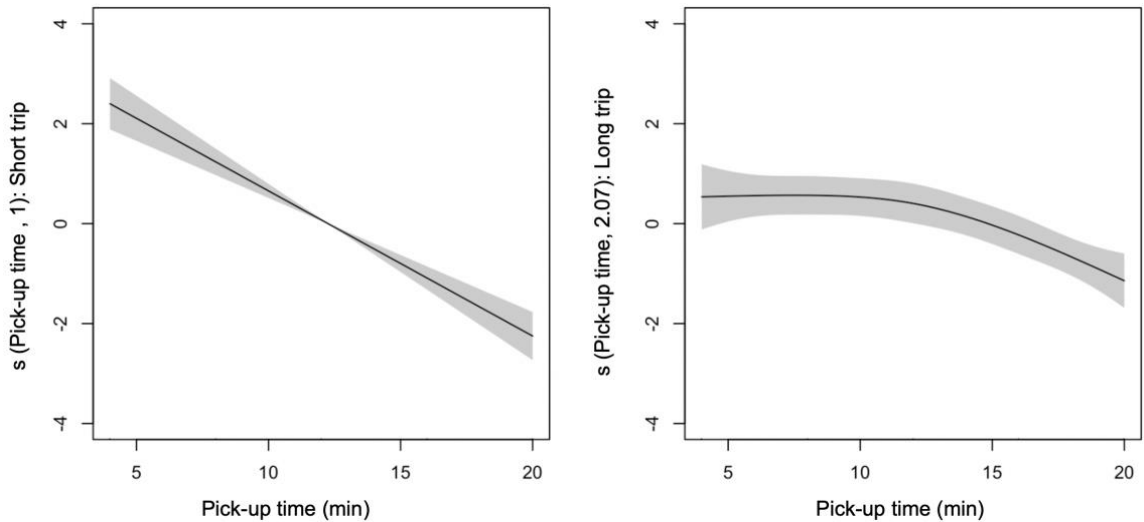

**Figure 3.** Nonlinear effects: pick-up time and long trip in GAMM model

(2) Linear terms

Surge price is positively associated with response to a request. It's reasonable since surge price means higher trip earnings. The result is also consistent with other studies (Ashkrof et al., 2021; Xu et al., 2018). However, surge pricing may also have downsides: one study mentioned that surge pricing may result in more rejected trips without surge prices because drivers were on the way to the surge pricing areas (Ashkrof et al., 2020).

Shared trip has a negative relationship with response to a request, which is consistent with a recent study (Morris et al., 2020). Shared trips might indicate lower trip fares, frequent pickups and drop-offs, and potentially complicated interactions between passengers (Morris et al., 2020). Another survey also discovered that drivers are generally dissatisfied with UberPool (Campbell, 2018).

Long trip (trip over 45 minutes) is significant and positively correlated with trip request acceptance. The result means ridehailing drivers are more likely to accept long trips than short ones. This is intuitive since overall, long trips mean a higher fare and less time lost on pickups and drop-offs (Ashkrof et al., 2020). However, some drivers mentioned a mixed feeling about long rides: a long trip risks leaving the driver in "the middle of nowhere" with no prospects for a return trip (Jones, 2015; Uber drivers' forum, 2015).

Passenger rating demonstrates a positive relationship with trip request acceptance, however, it the parameter is statistically significant only at 10% significance level. Generally, it is expected that ridehailing drivers may favor trips with higher passenger ratings. A few respondents in our survey mentioned in the comment area that they would not take passengers with ratings lower than a certain threshold (e.g., 4.5). This echoes existing evidence (Bowman, 2019; RideGuru, 2018). Drivers dislike low passenger ratings because such passengers might be late for the ride, disrespectful to the driver, and neglecting road safety (thus ask the driver to pick up/drop off in the middle of the road), among other reasons (Helling, 2021). However, when a spline term for passenger ratings was included in our model, it failed to indicate any threshold or non-linear effects.

Finally, average waiting time for a trip is negatively associated with trip request acceptance but is non-significant.

*Socio-demographics and employment characteristics*

The parameter estimation results of socio-demographics and employment status are found to be non-significant. This result is also confirmed by the results of a generalized likelihood ratio test (Fan & Jiang, 2007). This contrasts with another study, which finds gender, employment status and amount of years driving are significant predictors of trip request acceptance (Ashkrof et al., 2021). They found males, part-time drivers, and those who have better education more likely to accept the trip request. However, the coefficients of these variables in our study are directionally different and all of them are non-significant. The other study also used stated preference experiments but did not account for repeated measures for the same individual (Ashkrof et al., 2021). To compare with this study, we re-estimated our model without accounting for repeated measures (by omitting the random effects). When we did so, the socio-demographics and employment characteristics were found to be statistically significant. This may suggest individual-level characteristics (including socio-demographics and employment status) appear overly significant when repeated measures are not accounted for within the modeling framework.

**Table 6.** Modeling results (N = 1085). Cells include coefficient estimates, followed by standard errors in parentheses. Dependent variable is decision to accept (y=1) or decline (y=0) a trip request.

| Models | Baseline model | Full model |
|---|---|---|
| *Fixed effects: coefficients* | | |
| **Variable** | **Coefficient Estimate** (Standard Error) | |
| Intercept | -0.808 (1.932) | -0.371 (2.844) |
| Long trip | 1.127*** (0.186) | 1.259*** (0.224) |
| Average waiting time for a trip (minutes) | -0.008 (0.009) | -0.008 (0.011) |
| Surge price ($) | 0.060*** (0.021) | 0.082*** (0.025) |
| Star rating of passenger | 0.580* (0.316) | 0.668* (0.385) |
| Shared/pooled | -0.719*** (0.182) | -0.889*** (0.222) |
| What is your age? (reference: <40) | | |
| 40-64 | 0.433 (0.387) | 0.567 (0.695) |
| >=65 | 1.088 | 1.687 |

|  | (0.958) | (1.637) |
|---|---|---|
| Household size | -0.099 | -0.138 |
|  | (0.089) | (0.157) |
| Household income (reference: less than city median) | | |
|     Equal or higher to city median | 0.532 | 0.807 |
|  | (0.664) | (1.148) |
| Gender (reference: male) | | |
|     Female | 1.256 | 1.848 |
|  | (0.970) | (1.664) |
| What is the highest degree or level of school you have completed? (reference: less than high school) | | |
|     High school | -1.165 | -1.813 |
|  | (1.006) | (1.759) |
|     College | -1.360 | -1.918 |
|  | (1.005) | (1.754) |
|     Graduate school and above | -2.154* | -3.216 |
|  | (1.234) | (2.143) |
| What platforms do you drive for? | | |
|     Uber | -0.269 | -0.160 |
|  | (0.407) | (0.708) |
|     Lyft | -0.066 | -0.083 |
|  | (0.535) | (0.920) |
| Do you have another job? (reference: yes, full time) | | |
|     Yes, part time | 1.027 | 1.152 |
|  | (0.812) | (1.432) |
|     No other jobs | 0.994 | 1.212 |
|  | (0.663) | (1.179) |
| *Smooth terms: estimated degree of freedom (EDF)* | | |
| **Variable** | **EDF (Chi. Sq.)** | |
| s (Pick-up time): Long trip | - | 2.065*** |
|  |  | (18.145) |
| s (Pick-up time): Short trip | - | 1.001*** |
|  |  | (87.354) |
| s (Trip rejection rate) | - | 4.469 |
|  |  | (1.863) |
| **Random effect** | **EDF (Standard Dev.)** | |
| ID | 117.7*** | 130.362*** |
|  | (2.021) | (1.925) |
| *Model performance* | | |
| Deviance explained | 38.3% | 53.5% |
| Final Log likelihood | -370 | -278 |

Note: '***' refers to 0.01 significance level; '**' refers to 0.05 significance level; '*' refers to 0.1 significance level.

**Implications**

A lack of understanding on how drivers respond to a trip request impairs the development of effective interventions for promoting trip request acceptance rates. To prevent drivers from rejecting the trip requests, TNCs have adopted strategies like hiding most trip information from the drivers, giving up to a 15 minutes timeout, or deactivating driver accounts when they frequently reject too many trips (Dough, 2018; McFarland, 2016; RideGuru, 2020). These "stick" strategies have been criticized as rough and neglecting drivers' rights (McFarland, 2016). One critical downside of adopting such strategies is, it makes ridehailing drivers unhappy, and might hurt drivers' willingness to stay in the labor market (Bursztynsky, 2021; Morris et al., 2020).

In this section, possible positive "carrot" interventions for increasing trip request acceptance rates are discussed. Based on the model estimation results, trip features are the most significant factors that affect drivers' trip request acceptance decisions. Therefore, understanding drivers' willingness to accept (WTA) compensation for main trip features are warranted. Drivers' WTA compensation in this section are presented based on the GAMM estimation results. WTA equations used in our study are shown in equations (2~4). Other possible interventions are also briefly discussed.

*Willingness to accept (WTA) compensation*

*Shared trip*

Compared with solo trips, a shared trip decreases the total utility by $\beta_m$, in this case, 0.889. To compensate for the same amount of utility lost caused by the shared trip, the surge price needs to increase by $\Delta S$, in this case, $10.84. This result means that drivers need to be compensated by an extra $10 to accept a shared trip! This result paints a negative picture for the financial sustainability of pooling, especially since other research shows that travelers want to pay less, not more, for a pooled trip (Khaloei, Tu, Zou, & MacKenzie, 2022).

$$\Delta u_m = \beta_m \Delta M = -\beta_s \Delta s_m \qquad (2)$$

Where $\Delta u_m$ is the utility difference between a solo and a shared trip; $\beta_m$, $\beta_s$ are the coefficients of the shared/pooled ride and surge price, respectively; $\Delta M$ represents shifting from a solo to a shared ride. $\Delta s_m$ refers to the WTA to compensate for $\Delta u_m$.

*Shared trip*

Decreasing the passenger rating by 0.10 decreases the total utility by $\beta_p$, in this case, 0.067. The surge price needs to increase by $0.81 to compensate for the utility lost caused by passenger rating decrease. This result means that drivers would be willing to accept a trip with lower passenger ratings when compensated by an extra $0.81.

$$\Delta u_p = \beta_p \Delta P = -\beta_s \Delta s_p \qquad (3)$$

Where $\Delta u_p$ is the utility change when the passenger rating decreases by $\Delta P$; $\beta_p$ is the coefficient of passenger ratings; $\Delta P$ represents passenger rating change, in this case, 0.10. $\Delta s_p$ refers to the WTA needed to compensate for $\Delta u_p$.

*Pick-up time*

Since the relationship between pick-up time and WTA changes is nonlinear, WTA of pick-up time was calculated by simulation (Figure 4).

$$\Delta U_r = g(y) - g(y_{(\Delta R)}) = -\beta_s \Delta S_r \tag{4}$$

Where $\Delta U_r$ is a vector of utility changes when every element in the pick-up time vector decreases by $\Delta R$; $g(\cdot)$ is a logit function; $y$ means the vector of driver responses to a trip request of the collected data; $y_{(\Delta R)}$ represents the simulated vector of driver responses when pick-up time increases by $\Delta R$. In this case, $\Delta R$ equals 1 minute. $\Delta S_r$ refers to a vector of prices needed to compensate for $\Delta R$ with different pick-up times.

Simulation results suggest that compared to short trips, WTA compensation for long trips is generally lower. For a short trip, it needs an additional $3.60 to accept a trip if the pick-up time increases by 1 minute, regardless of the initial pick-up time. However, for a long trip, drivers only need compensation if the pick-up time goes beyond 8 minutes. For instance, when the pick-up time increases from 20 to 21 minutes, ridehailing drivers would need an extra $3 to keep the utility of the trip the same.

One thing to note is that although we investigated drivers' willingness to acceptance surge prices, surge price may not function entirely the same as higher trip fare. For example, some drivers mentioned surge pricing cannot be trusted since it only lasts for several minutes. They do not chase surge prices since it often disappears when they arrive at an area that moments before had a high surge price. A higher fare that is guaranteed for each trip may have a larger impact on promoting trip request acceptance. Future research may test the effect of higher trip fare on trip request acceptance.

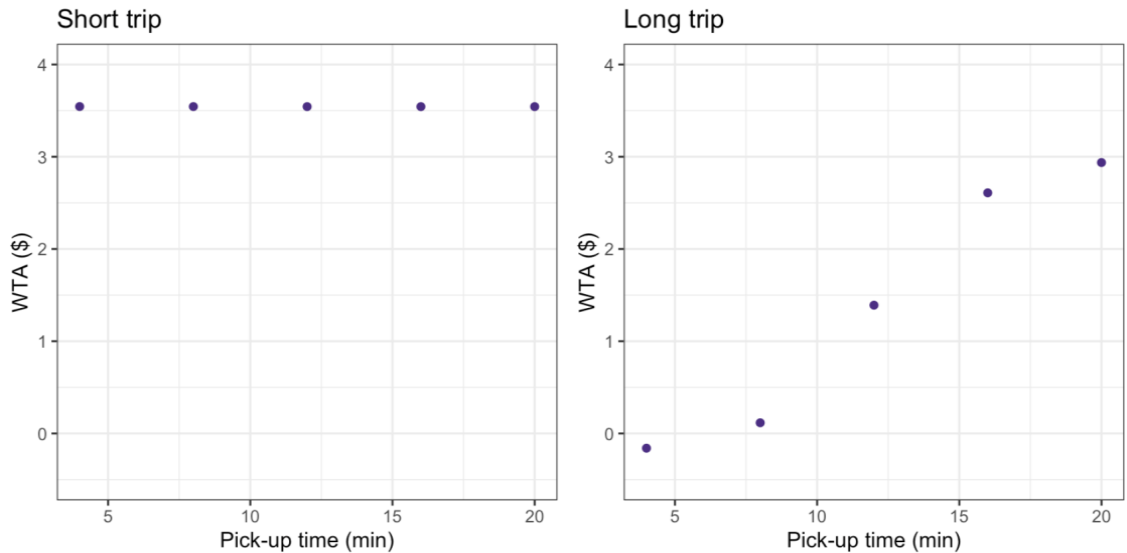

**Figure 4.** WTA compensation for pick-up time

*Other interventions*

Aside from monetary compensation, some other interventions could promote ridehailing drivers' trip request acceptance rates. This may include, better passenger training, more satisfying driver protection mechanisms, and more efficient algorithms to match passengers. Such interventions may improve passenger ratings, decrease pick-up times, and reduce drivers' cost of accepting shared ride trips and trips with lower passenger ratings.

**Conclusions**

Understanding trip request acceptance of ridehailing drivers is significant for improving ridehailing service efficiency, promoting equal access, and easing negative environmental impacts. This study explored the non-linear relationship between trip request acceptance and trip features, socio-demographics and employment characteristics using a generalized additive mixed model (GAMM).

The model results confirm that, firstly, nonlinear effects between trip features and trip request acceptance are necessary to explore, and secondly, it is important to account for repeated measures by including random effects in the model.

Among all the variables, trip features have the most significant effects. Results suggested that longer pick-up time is negatively associated with trip request acceptance. The relationship between pick-up time and trip request acceptance is linear when the trip is short (not over 45 minutes), while nonlinear otherwise. When the trip is short, drivers are more likely to accept trips with shorter pick-up times, and less likely to accept trips with longer pick-up times compared to when the trip is long. When the trip is long, the pick-up time has smaller positive effect on the trip request acceptance when the trip is less than 12 minutes. Once the pick-up time goes beyond 12 minutes, the trip request acceptance drops rapidly. It was also found that long trips (over 45 minutes) are positively associated with trip request acceptance. With an increase in surge price, drivers are more likely to accept trip requests. Furthermore, passenger ratings are positively associated with trip request acceptance. However, drivers are less likely to accept shared trips than solo ones.

Drivers' socio-demographic and employment status impacts were statistically non-significant in our model. However, their impacts became significant when repeated measure effects were no longer controlled. This points to the importance of controlling for repeated measures and highlights an important limitation of some of the extant literature that found socio-demographic predictors to be significant, but failed to control for repeated measures.

As for corresponding policy interventions to promote trip request acceptance, we concluded that the most effective approach might be adjusting trip features (e.g., increase surge price, improve matching algorithms to decrease pick-up time) or compensating for undesired trip features. However, designing policies targeting drivers with different socio-demographics may not work, since their impacts on trip request acceptance are non-significant after accounting for repeated measures.

**Future work**

There are several future directions in which our work can be improved and extended. First, the sample in our study only comes from the airport waiting area. The ridehailing drivers who were allowed to be at the SeaTac airport waiting area had two characteristics: (1) they have a King County for-hire driver's permit; (2) they have an electric or hybrid car. Those whose permit number was expired or those who had conventional gasoline cars were therefore excluded from the sampling frame. Future studies might consider including drivers from other channels (e.g., hailing rides) to provide a more representative view of ridehailing drivers. However, our experience was that ridehailing drivers are a notoriously difficult community to reach and recruit for survey research. The airport waiting area was the only recruitment channel that provided a high response rate and an acceptable cost per response[2].

Second, certain factors were not accommodated within the modeling frameworks that may influence drivers' response to a trip request. For example, the City of Seattle has a higher pay rate than surrounding cities such as Tacoma and Renton, thus drivers might prefer trips within the city of Seattle. Another example is that the COVID-19 pandemic may have an impact on ridehailing drivers' trip acceptance behaviors (Shokoohyar et al., 2022; Shokouhyar et al., 2021). In addition, it may be worthwhile to explore the relationship between trip length and trip acceptance in a more granular scale (we use the coarse measure of *short vs long trip* in this study). Future efforts should identify and consider such factors to have a better understanding of the relationship between trip features and drivers' behavior.

---

[2] $23 for driver compensation ($15 + 55.5% overhead)

Third, future investigators may want to explore systematically the reasons that drivers prefer solo over pooled trips. Anecdotally, in an open response section of the survey, drivers offered several reasons they dislike pooled trips. For one, pooled trips are not as profitable as the solo ones. For another, pooled trips are more work than solo ones since the driver needs to pick up and drop off more than one passenger. Finally, some drivers mentioned that passengers give them lower ratings because the ride experience is not as good on a solo trip; passengers pay a lower price in a pooled trip but seem to have the same expectations as for a solo trip. Understanding the reasons drivers dislike pooled trips would help to inform company and government policies that might help to promote pooling and realize its congestion, energy, and environmental benefits.

Fourth, future studies might consider collecting revealed preference data (e.g., trip records) from TNC companies to supplement the stated preference data in our study.

Despite these limitations, this study provides critical insights on ridehailing drivers' behavior towards passenger trip requests. Findings from this research would assist the transportation planners and policymakers to create policy interventions that would focus on improving ridehailing driver behaviors and develop more efficient transportation systems.

**Contribution statement**

**Yuanjie (Tukey) Tu:** Conceptualization, Survey design, Data collection, Methodology, Formal analysis, Software, Writing - Original draft preparation; **Moein Khaloei:** Survey design, Data collection; **Nazmul Arefin Khan:** Writing – Review & Editing, Supervision; **Don MacKenzie:** Conceptualization, Survey design, Methodology, Writing – Review & Editing, Supervision, Project administration, Funding acquisition. All authors read, commented, and approved the final version of the manuscript.

**Acknowledgement**

The work was supported by the U.S. Department of Energy (DOE) Vehicle Technologies Office (VTO) under the Systems and Modeling for Accelerated Research in Transportation (SMART) Mobility Laboratory Consortium, an initiative of the Energy Efficient Mobility Systems (EEMS) Program. The submitted manuscript has been created by University of Washington and the UChicago Argonne, LLC, Operator of Argonne National Laboratory (Argonne). Argonne, a U.S. Department of Energy Office of Science laboratory, is operated under Contract No. DE-AC02-06CH11357. The U.S. Government retains for itself, and others acting on its behalf, a paid-up nonexclusive, irrevocable worldwide license in said article to reproduce, prepare derivative works, distribute copies to the public, and perform publicly and display publicly, by or on behalf of the Government.

**Reference**

Abdallah, W., Goergen, M., & O'Sullivan, N. (2015). Endogeneity: How Failure to Correct for it can Cause Wrong Inferences and Some Remedies. *British Journal of Management*, *26*(4), 791–804. https://doi.org/10.1111/1467-8551.12113

Ashkrof, P., Correia, G. H. de A., Cats, O., & van Arem, B. (2020). Understanding ride-sourcing drivers' behaviour and preferences: Insights from focus groups analysis. *Research in Transportation Business & Management*, *37*, 100516. https://doi.org/10.1016/j.rtbm.2020.100516

Ashkrof, P., de Almeida Correia, C. H., Cats, O., & Van Arem, B. (2021). *Ride Acceptance Behaviour of Ride-sourcing Drivers*. 25. https://arxiv.org/abs/2107.07864

Balk, G. (2020, September 25). *Seattle's median household income soars past $100,000—But wealth doesn't reach all*. The Seattle Times. https://www.seattletimes.com/seattle-news/data/seattles-median-income-soars-past-100000-but-wealth-doesnt-reach-all/

Barnes, S. J., Guo, Y., & Borgo, R. (2020). Sharing the air: Transient impacts of ride-hailing introduction on pollution in China. *Transportation Research Part D: Transport and Environment*, *86*, 102434. https://doi.org/10.1016/j.trd.2020.102434

Bokányi, E., & Hannák, A. (2020). Understanding Inequalities in Ride-Hailing Services Through Simulations. *Scientific Reports*, *10*(1), 6500. https://doi.org/10.1038/s41598-020-63171-9

Bowman, C. (2019). *I drive for Uber and Lyft—Here's what your driver thinks about you based on your rating*. Business Insider. https://www.businessinsider.com/uber-lyft-rating-passenger-meaning-2019-11

Brown, A. E. (2020). Who and where rideshares? Rideshare travel and use in Los Angeles. *Transportation Research Part A: Policy and Practice*, *136*, 120–134. https://doi.org/10.1016/j.tra.2020.04.001

Bursztynsky, J. (2021, July 4). *Why many Uber and Lyft drivers aren't coming back*. CNBC. https://www.cnbc.com/2021/07/04/why-many-uber-and-lyft-drivers-arent-coming-back.html

Campbell, H. (2018). *The Rideshare Guy—2018 Uber and Lyft Driver Survey*. Google Docs. https://docs.google.com/document/d/1g8pz00OnCb2mFj_97548nJAj4HfluExUEgVb45HwDrE/edit?usp=sharing&usp=embed_facebook

Diao, M., Kong, H., & Zhao, J. (2021). Impacts of transportation network companies on urban mobility. *Nature Sustainability*, *4*(6), 494–500. https://doi.org/10.1038/s41893-020-00678-z

Dough. (2018, February 7). Fired from Uber: Why Drivers Get Deactivated, & How to Get Reactivated. *Ridesharing Driver*. https://www.ridesharingdriver.com/fired-uber-drivers-get-deactivated-and-reactivated/

Erhardt, G. D., Mucci, R. A., Cooper, D., Sana, B., Chen, M., & Castiglione, J. (2021). Do transportation network companies increase or decrease transit ridership? Empirical evidence from San Francisco. *Transportation*. https://doi.org/10.1007/s11116-021-10178-4

Fan, J., & Jiang, J. (2007). Nonparametric inference with generalized likelihood ratio tests. *TEST*, *16*(3), 409–444. https://doi.org/10.1007/s11749-007-0080-8

Ge, Y., Knittel, C. R., MacKenzie, D., & Zoepf, S. (2020). Racial discrimination in transportation network companies. *Journal of Public Economics*, *190*, 104205. https://doi.org/10.1016/j.jpubeco.2020.104205

Graehler, M., Mucci, R., & Erhardt, G. D. (2019). Understanding the Recent Transit Ridership Decline in Major US Cities: Service Cuts or Emerging Modes? *Transportation Research Board Annual Meeting 98th*, 19.

Hassanpour, A., Bigazzi, A., & MacKenzie, D. (2021). Equity of access to Uber's wheelchair accessible service. *Computers, Environment and Urban Systems*, *89*, 101688. https://doi.org/10.1016/j.compenvurbsys.2021.101688

Helling, B. (2021, July 26). *Uber Rider Ratings: Everything You Need To Know & Succeed | Ridester.com*. https://www.ridester.com/uber-rider-ratings/

Hughes, R., & MacKenzie, D. (2016). Transportation network company wait times in Greater Seattle, and relationship to socioeconomic indicators. *Journal of Transport Geography*, *56*, 36-44.

Jones, L. (2015, July 17). *How Your Uber Driver Really Feels About Long Rides*. LAist. https://laist.com/news/with-the-promise-of-a

Khaloei, M., Tu, Y., Zou, T., & MacKenzie, D. (2022). *How will eliminating drivers from autonomous ridehailing services affect pooled ridehailing?*

McFarland, M. (2016, July 28). *How Uber punishes drivers who refuse to use UberPool*. CNNMoney. https://money.cnn.com/2016/07/28/technology/uber-uberpool-timeouts/index.html

Morris, E. A., Zhou, Y., Brown, A. E., Khan, S. M., Derochers, J. L., Campbell, H., Pratt, A. N., & Chowdhury, M. (2020). Are drivers cool with pool? Driver attitudes towards the shared TNC services UberPool and Lyft Shared. *Transport Policy*, *94*, 123–138. https://doi.org/10.1016/j.tranpol.2020.04.019

Payyanadan, R. P., & Lee, J. D. (2018). Understanding the ridesharing needs of older adults. *Travel Behaviour and Society*, *13*, 155–164. https://doi.org/10.1016/j.tbs.2018.08.002

RideGuru. (2018). *As a rider, do you really care about your Uber or Lyft score? - RideGuru*. https://ride.guru/lounge/p/as-a-rider-do-you-really-care-about-your-uber-or-lyft-score

RideGuru. (2020). *Uber still punishes for declining January 2020—RideGuru*. https://ride.guru/lounge/p/uber-still-punishes-for-declining-january-2020

Shokoohyar, S. (2018). Ride-sharing platforms from drivers' perspective: Evidence from Uber and Lyft drivers. *International Journal of Data and Network Science*, 89–98. https://doi.org/10.5267/j.ijdns.2018.10.001

Shokoohyar, S., Jafari Gorizi, A., Ghomi, V., Liang, W., & Kim, H. J. (2022). Sustainable Transportation in Practice: A Systematic Quantitative Review of Case Studies. *Sustainability*, *14*(5), 2617. https://doi.org/10.3390/su14052617

Shokoohyar, S., Sobhani, A., & Ramezanpour Nargesi, S. R. (2020). On the determinants of Uber accessibility and its spatial distribution: Evidence from Uber in Philadelphia. *WIREs Data Mining and Knowledge Discovery*, *10*(4), e1362. https://doi.org/10.1002/widm.1362

Shokoohyar, S., Sobhani, A., & Sobhani, A. (2020). Impacts of trip characteristics and weather condition on ride-sourcing network: Evidence from Uber and Lyft. *Research in Transportation Economics*, *80*, 100820. https://doi.org/10.1016/j.retrec.2020.100820

Shokouhyar, S., Shokoohyar, S., Sobhani, A., & Gorizi, A. J. (2021). Shared mobility in post-COVID era: New challenges and opportunities. *Sustainable Cities and Society*, *67*, 102714. https://doi.org/10.1016/j.scs.2021.102714

Tu, Y., Jabbari, P., & Khaloei, M. (2021). *CERC and SMART Report Year 4.docx*. https://works.bepress.com/yuanjie-tu/1/

Uber drivers' forum. (2015). *Do you all prefer long trips?* Uber Drivers Forum. https://www.uberpeople.net/threads/do-you-all-prefer-long-trips.44415/

Wang, M., & Mu, L. (2018). Spatial disparities of Uber accessibility: An exploratory analysis in Atlanta, USA. *Computers, Environment and Urban Systems*, *67*, 169–175. https://doi.org/10.1016/j.compenvurbsys.2017.09.003

Ward, J. W., Michalek, J. J., Azevedo, I. L., Samaras, C., & Ferreira, P. (2019). Effects of on-demand ridesourcing on vehicle ownership, fuel consumption, vehicle miles traveled, and emissions per capita in U.S. States. *Transportation Research Part C: Emerging Technologies*, *108*, 289–301. https://doi.org/10.1016/j.trc.2019.07.026

Wentrup, R., Nakamura, H. R., & Ström, P. (2018). Uberization in Paris – the issue of trust between a digital platform and digital workers. *Critical Perspectives on International Business*, *15*(1), 20–41. https://doi.org/10.1108/cpoib-03-2018-0033

Wood, S. (2021). *mgcv: Mixed GAM Computation Vehicle with Automatic Smoothness Estimation* (1.8-38). https://CRAN.R-project.org/package=mgcv

Wood, S. N. (2017). *Generalized additive models: An introduction with R* (Second edition). CRC Press/Taylor & Francis Group.

Wu, X., & MacKenzie, D. (2021). The evolution, usage and trip patterns of taxis & ridesourcing services: Evidence from 2001, 2009 & 2017 US National Household Travel Survey. *Transportation*. https://doi.org/10.1007/s11116-021-10177-5

Xu, K., Sun, L., Liu, J., & Wang, H. (2018). An empirical investigation of taxi driver response behavior to ride-hailing requests: A spatio-temporal perspective. *PLOS ONE*, *13*(6), e0198605. https://doi.org/10.1371/journal.pone.0198605